# Manipulating Optical Nonlinearities of Molecular Polaritons by Delocalization

Bo Xiang[2]†, Raphael F. Ribeiro[1]†, Yingmin Li[2], Adam D. Dunkelberger[3], Blake B. Simpkins[3], Joel Yuen-Zhou[1], Wei Xiong[1,2]*

[1]Department of Chemistry and Biochemistry, University of California, San Diego, La Jolla, CA, 92093

[2]Materials Science and Engineering Program, University of California, San Diego, La Jolla, CA, 92093

[3]Chemistry Division, Naval Research Laboratory, Washington, District of Columbia, 20375

**Optical nonlinearities are key resources in the contemporary photonics toolbox, relevant to quantum gate operations and all-optical switches. Chemical modification is often employed to control the nonlinear response of materials at the microscopic level, but on-the-fly manipulation of such response is challenging. Tunability of optical nonlinearities in the mid-IR is even less developed, hindering its applications in chemical sensing or IR photonic circuitry. Here, we report control of vibrational polariton coherent nonlinearities by manipulation of macroscopic parameters such as cavity longitudinal length or molecular concentration. Further 2D IR investigations reveal that nonlinear dephasing provides the dominant source of the observed ultrafast polariton nonlinearities. The reported phenomena originate from the nonlinear macroscopic polarization stemming from strong coupling between microscopic molecular excitations and a macroscopic photonic cavity mode.**

**INTRODUCTION.**

Molecular polaritons are hybrid quasiparticles resulting from strong coupling between molecular excitations(*1–4*) and confined electromagnetic degrees of freedom, e.g. Fabry-Pérot (FP) microcavity modes. Heuristically, polaritons arise when a cavity photon mode interconnects the microscopic molecular degrees of freedom, rendering their wavefunction coherently delocalized across a macroscopic length scale. This interplay between microscopic and macroscopic characteristics is at the heart of the versatility of polaritons. Much of the recent research on molecular polaritons has focused on the modification of molecular properties through the interaction with an optical cavity(*5–14*). Conversely, control of polariton optical response via manipulation of the macroscopic parameters which define the electromagnetic modes (e.g., cavity length) has received far less attention(*15*) (Fig.1a) and, in fact, we are not aware of any investigations of these effects in the nonlinear photonic regime.

In this work, we first reveal that ultrafast molecular polariton nonlinearities can be conveniently manipulated by controlling the size of the optical cavity and the molecular concentration. The ease and adaptability of such operations make them applicable to in-situ control of optical nonlinearities for photonic circuitry and quantum information processing applications(*16, 17*). The microscopic origin of the distinctive polariton optical signals is subsequently investigated via analysis of the vibrational polariton 2D IR spectral dynamics. In contrast to generic molecular nonlinearities which tend to be modulated by excited-state population, we find that nonlinear dephasing is the main contributor to the coherent polariton optical response. We conclude by commenting on the potential implications of this research to the deployment of previously undeveloped infrared devices.

†These authors contributed equally.

# RESULTS.

***Polariton bleach dependence on cavity longitudinal length.*** To investigate the optical nonlinearities of vibrational polaritons, we measure their third-order nonlinear susceptibilities by femtosecond IR pump-probe spectroscopy. The hybrid light-matter system consists of a FP microcavity filled with an ensemble of asymmetric carbonyl stretch modes originating from $W(CO)_6$ molecules in a hexane solution(*14*). At zero waiting time, when IR pump and probe pulses overlap, we see a notable reduction in the intensity of the polariton transmission (Figs. 1b). This reduction gives rise to absorptive features in the differential transmission spectra (Figs. 1c). Qualitatively, the observed behavior resembles the well-known "photon blockade effect" in the single-emitter quantum regime of the Jaynes-Cummings model(*18–20*): when a photon excites the emitter-cavity system, the latter is blocked from further interactions with incoming photons of the same frequency. Here, we observe a similar effect, although it is in the ensemble regime and the mechanism for reduced transmission is quite different from the single oscillator case. We hereafter term this phenomenon "polariton bleach". Note this effect is not a trivial coherent artifact from the overlap between the pump and probe pulses. The IR pulse duration is ~100 fs, while the bleach lasts for less than 5 ps (which is also the approximate lifetime of the cavity photon), indicating the effect is intrinsic to the lifetime of the cavity polaritons. The polariton bleach concept provides a foundation for developing mid-IR photonic devices in the ensemble regime of light-matter strong coupling.

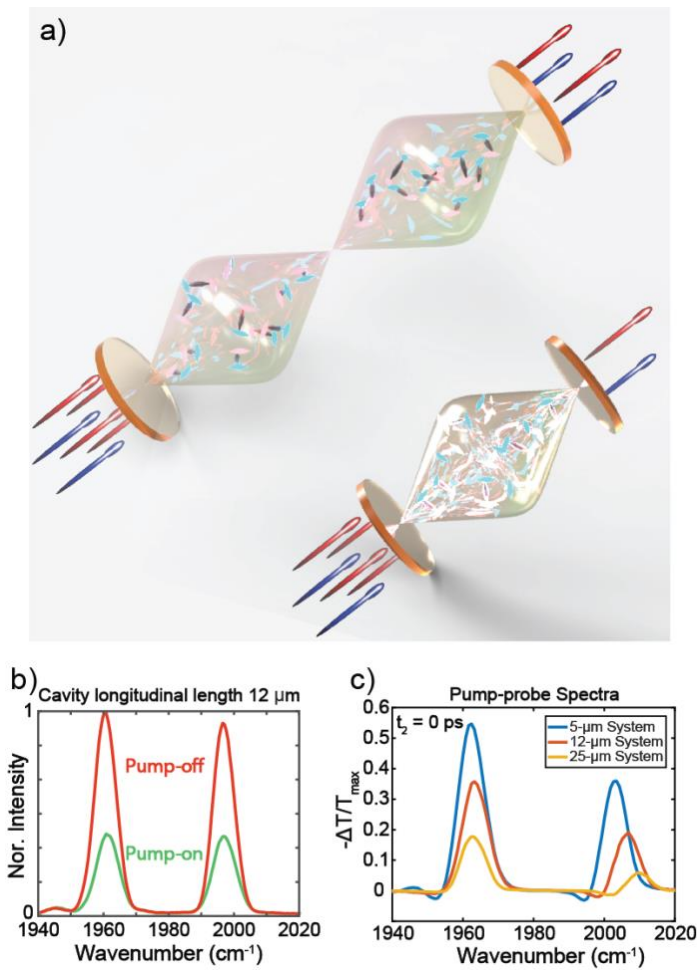

*Figure 1. Polariton bleach effect at pum-probe delay time, $t_2$ = 0 ps. (a) The central concept for optical nonlinearity manipulation via changes in the cavity thickness is that by decreasing the cavity longitudinal length, and consequently the mode volume, the pump-induced polariton density is increased along with the corresponding optical nonlinearity. The wave-like feature represents the microcavity standing-waves in resonance with the molecular modes. The resonant cavity mode decreases when the cavity length is reduced (e.g., in our experiment, the 10th and 5th order cavity modes are nearly-resonant with the molecular polarization when the cavity longitudinal length is 25 and 12 μm, respectively). The red and blue rays entering and leaving the FP cavities provide a pictorial representation of input and output (detected) electromagnetic fields, and do not imply any focusing or laser profile. Further description of the experimental setup is given in SI Secs. S1.1 and S1.2. (b) The intensity of the bleach is presented by comparing the transmission intensity when pump is on (green) and off (red) in the 12 μm cavity. (c) Pump-probe spectra showing the polariton bleach with various cavity longitudinal lengths (5, 12 and 25 μm). It is found that the polariton bleach (reduction of transmission) becomes weaker as the cavity longitudinal length increases. All pump-probe spectra are normalized to the maximum intensity of the pump-off transmission spectra. All measurements were performed with a fixed molecular concentration, Rabi splitting and pump input (~10 mJ/cm$^2$).*

The polariton bleach shows peculiar dependence on the cavity thickness: it is dramatically *stronger* when the cavity longitudinal length is decreased. In fact, as shown in Figs. 1c, when the cavity thickness decreases from 25 to 5 μm, the magnitude of the polariton bleach increases. This observation leads to the remarkable conclusion that substantially enhanced nonlinearities are induced on molecular vibrational polaritons by merely changing the longitudinal length of the optical microcavity. Importantly, the concentration of $W(CO)_6$ is kept constant so that polariton resonance frequencies detected with linear transmission measurements are independent of the cavity thickness (the Rabi splitting is fixed within 2 cm$^{-1}$). The variation in Rabi splitting is due to the different lifetimes associated to each cavity thickness(*21*). From a molecular spectroscopy standpoint, this is an unexpected and important result given that a sample of independent emitters in the *weak* light-matter coupling regime features nonlinear optical signals (e.g., 3rd order to 1st order nonlinear transmission ratios, $\zeta = \Delta T^{(3)}/T^{(1)}$) that remain unchanged upon scaling of system size (see Supplementary Materials S.1.4), *i.e.*, molecular nonlinearities are typically intensive properties of the system(*22*).

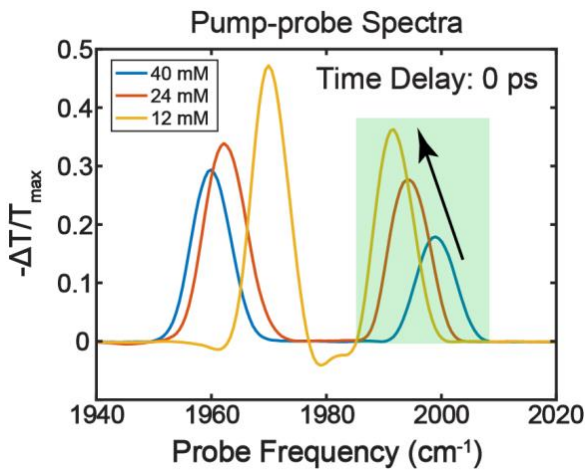

*Figure 2. Polariton bleach effect at pump probe delay time ($t_2$) of 0 ps (optical nonlinearity) as a function of concentration. From the pump-probe spectra, the bleach signal (normalized to the maximal linear transmission peak intensity) increases as the molecular concentration decreases.*

***Polariton bleach dependence on molecular concentration.*** In addition to a cavity-length dependence, we also observed that the magnitude of the polariton bleach can be tuned by changing the concentration of the molecular oscillators in a fixed-thickness cavity. Here we show pump-probe spectra performed with the 25 μm cavity at zero waiting time as a function of molecular concentration (Fig. 2). We focus on the transient signal at $\omega_{probe} = \omega_{UP}$ (Fig. 2) to avoid complications of the spectral peak at $\omega_{probe} = \omega_{LP}$ stemming from overlap with the 1→ 2 dark mode transition(*6, 13, 14*). The upper polariton (UP) peak intensities (faded green region in Fig. 2) show an inverse dependence on the molecular absorber concentration. Again, this is a striking effect from the perspective of conventional molecular spectroscopy in the weak light-matter coupling regime, where $\zeta$ typically remains constant with respect to molecular concentration (i.e., the molecules act as independent quantum oscillators). Importantly, these results *also* suggest the macroscopic dependence of the nonlinear optical response on cavity length *cannot* be solely ascribed to the variations of intracavity electromagnetic energy density because intracavity energy density is not affected by molecular concentration. We note that the choices of concentration and cavity thickness are limited by the requirement of strong coupling. Assuming normal incidence, there is a discrete set of cavity lengths approximately equal to $\frac{n\lambda}{2}$ (where *n* is the order of the mode and $\lambda$ is the molecular resonant wavelength) which makes 5, 12 and 25 μm the only readily available thicknesses for use in this experiment (due to practical spacer thickness options). Similarly, the 40-mM solution is nearly saturated, and 12 mM is the lowest concentration we achieve before the onset of weak-coupling, so we have explored the entire available cavity thickness and concentration range.

The nonlinear signal dependence on the cavity thickness (Figures 1c) and molecular concentration (Figure 2) show that an entirely new scaling of nonlinear interactions with size and concentration occurs in the molecular strong coupling regime. These properties can be qualitatively understood as consequences of the ratio between generated polaritonic excitations to the total number of molecular states comprising the polariton wave functions. When either molecular concentration or cavity thickness is changed, the absolute *number* of polariton transitions accessed by the IR pulses is approximately unchanged (determined by the number of IR photons entering the cavity). However, the number of molecular states composing polariton wavefunctions is increased when the cavity thickness or molecular concentration increases, thus leading to a reduced probability for any given molecule to be detected in an excited-state, and therefore resulting in the weakening of nonlinearities. Quantitatively, the observed *inverse* dependence of the polariton nonlinear response with respect to cavity thickness and molecular concentration is mediated by intra and intermolecular anharmonicity (more details can be found in the Supplementary Materials Sec. S3.1). These interactions lead to polariton nonlinear response proportional to the ratio of the pump-induced polariton density $\rho_{pol}(E_{pump})$ to the molecular density ρ. To see this, note that the nonlinear signal can be written in terms of a sum over all pump-driven polaritons, which scales with $\sum_{\mathbf{k}} \frac{|E_{pump}(\mathbf{k})|^2}{N(\mathbf{k})} \propto \frac{\rho_{pol}(E_{pump})}{\rho}$, where the in-plane cavity wave-vectors are labeled by $\mathbf{k}$, and the average number of molecules composing the corresponding polariton is $N(\mathbf{k})$. The 1/*N* factor arises from the dilution of molecular anharmonicity by the polariton volume (see Supplementary Materials Sec. S3.1).

The preceding argument clarifies the observed scaling of polariton nonlinearity with system size and molecular concentration. However, it does not reveal the *microscopic* nonlinear dynamical process which gives rise to the distinctive polariton bleach lineshape. In the next section, we further explore the microscopic mechanism responsible for the polariton bleach effect.

***Origin of Polariton Bleach and Rabi Oscillation Dynamics Revealed by 2D IR.*** Previous research on inorganic semiconductor exciton polaritons has shown that Coulomb scattering by charge carriers leads to similar phenomena to the polariton bleach described above (*23–25*). The same mechanism is *unlikely* to apply for localized molecular vibrations (or Frenkel excitons which form organic exciton-polaritons under strong coupling), since local molecular modes typically only interact weakly with each other via dipole-dipole couplings. These considerations motivate our investigation of the physical mechanism corresponding to the polariton bleach using 2D IR (Fig. 3a) spectroscopy. 2D IR measures the third-order nonlinear optical response function of a material, providing detailed spectral and dynamical features(*14*, *26–31*) hidden from pump-probe spectra. Specifically, 2D IR can state-selectively excite and probe particular transitions, a feature which is critical to discern the origin of nonlinearities in polaritons and dark modes. In Fig. 3d, we show the 2D IR spectrum of vibrational polaritons (cavity thickness = 25 μm) at $t_2$ = 0 ps (the time delay between the second and the third pulse; see more details of 2D-IR setup in Supplementary Materials Sec. S1.2). Four absorptive peaks dominate the spectra: two lie along the diagonal while the remaining correspond to cross-peaks. The polariton 2D IR spectra obtained with the 12 μm-cavity display qualitatively similar features (Fig. S5a). Spectral cuts of 2D IR that correspond to the nonlinear response obtained by pumping specific polaritons (UP or LP) explicitly show that all peaks have absorptive lineshapes. Furthermore, comparison of the 12 and 25 μm cavity-polariton spectral cuts (Fig. S5e and S5f) demonstrates agreement with the pump-probe results – there are stronger nonlinearities in the thinner cavity. We note that, due to heterogeneity and polariton relaxation, dark modes may also be weakly excited, and induce similar absorptive features (*14*). Overall, the observed 2D IR spectral features and the corresponding cavity longitudinal length dependence are consistent with the polariton bleach observed in the pump-probe data.

Additional insights on the nonlinear dynamics giving rise to polariton bleach can be obtained by examining the corresponding spectral lineshape evolution. Specifically, both the pump-probe and 2D IR signals at $t_2 = 0$ ps (Figs. 3b and 3d) show different spectral lineshapes from those measured at $t_2 = 25$ ps (Figs. 3c and 3e). At 25 ps, the peaks located at the probe frequency $\omega_3 = \omega_{LP}$ are much stronger than those at $\omega_3 = \omega_{UP}$, and the signal centered around $\omega_3 = \omega_{UP}$ evolves into a derivative shape. As explained in previous work(*13,*

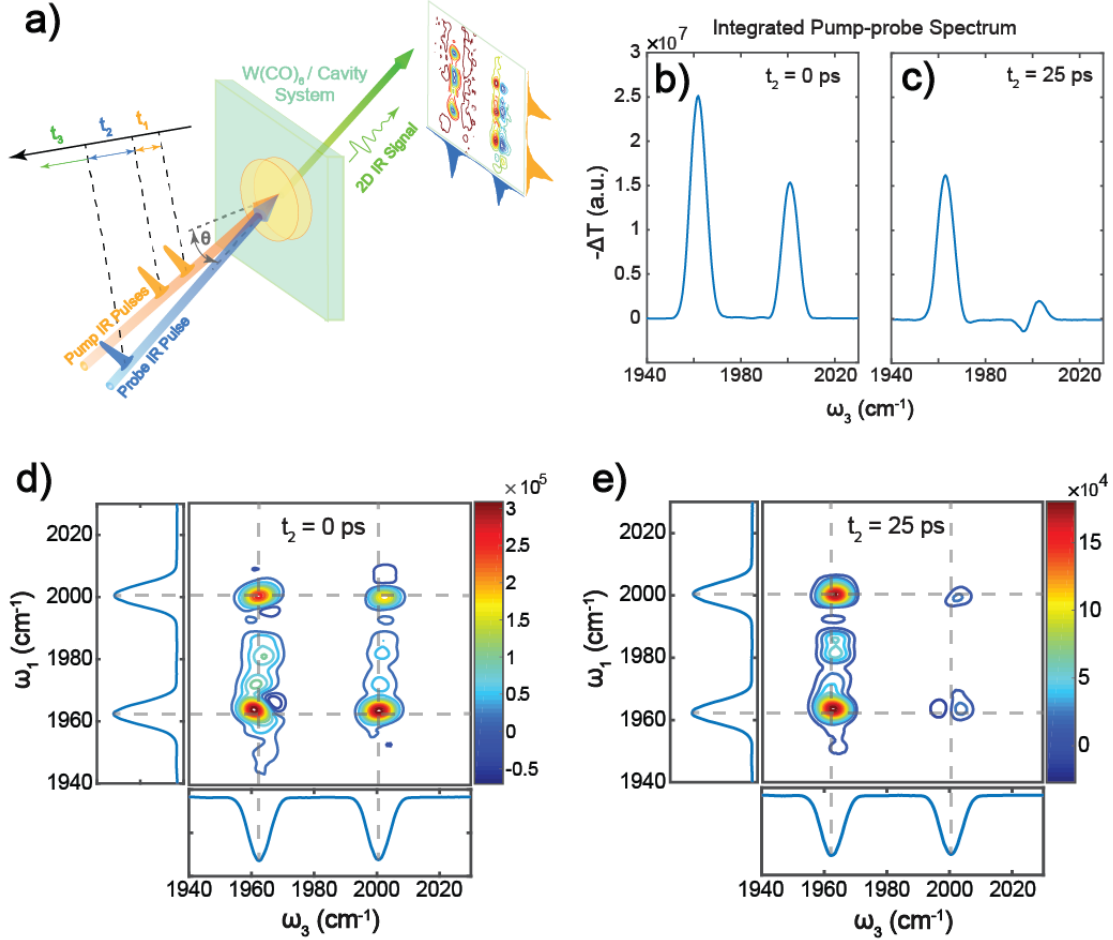

**Figure 3. 2D IR spectroscopy of molecular vibrational polaritons (25 μm optical cavity).** *(a) Illustration of 2D IR experimental setup (see Sec. S1.2 for details). a.u., arbitrary units. In (b) and (c), we show pump-probe spectra at early and late time delays (in comparison to the cavity photon lifetime which is approximately 5 ps). In (d) and (e), 2D IR spectra of molecular vibrational polaritons are provided at early and late time delays. Strong absorptive lineshapes are observed at early time delays.*

*14*), these features may be viewed to arise from ground-state population bleach, or equivalently, by the existence of transient excited reservoir population. The large absorptive peaks at $\omega_3 = \omega_{LP}$ arise primarily from the interference of the LP transition with the excited state absorption of dark reservoir population. The dispersive lineshape at $\omega_3 = \omega_{UP}$ is mainly a result of the Rabi splitting contraction induced by the bleaching of molecular absorbers. In contrast, the polariton bleach at $t_2 = 0$ ps has a qualitatively distinct lineshape with the feature at $\omega_3 = \omega_{UP}$ being stronger and purely absorptive (see Figs. 3b and 3c). This indicates that vibrational-polariton dynamics is significantly different at ultrafast (shorter than the polariton lifetime ~5 ps) from the one at late times (e.g. 25 ps).

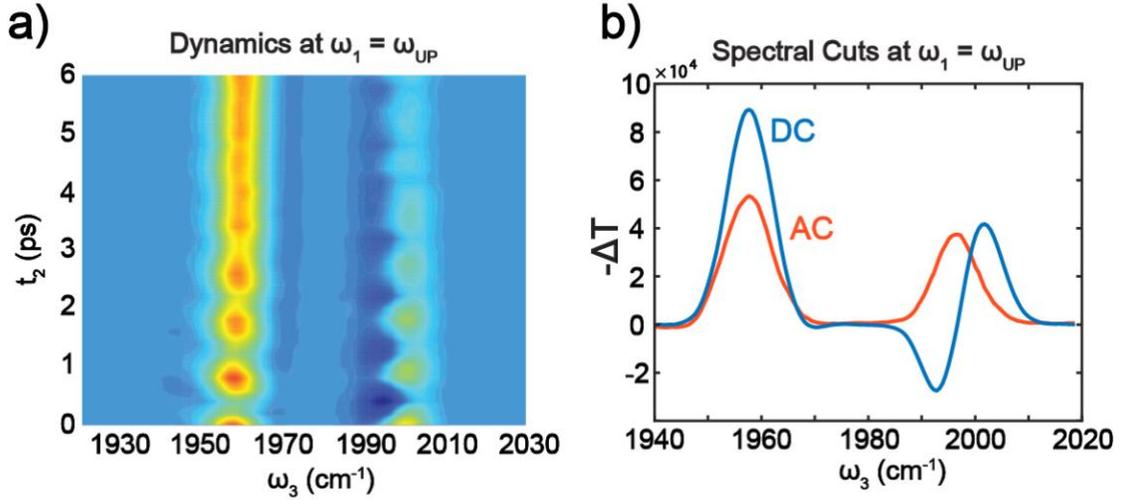

*Figure 4. 3D fast Fourier Transform results: (a) Time-dependent nonlinear optical response (differential transmission) showing Rabi oscillations (AC) of the 2D IR spectral cuts at $\omega_1=\omega_{UP}$ between 0 and 6 ps, on top of a decaying dynamics (DC). (b) AC and DC spectral cuts of differential transmission at $\omega_1=\omega_{UP}$, showing a purely absorptive lineshape in the AC spectrum and a dispersive lineshape in the DC spectrum at $t_2$=0.8 ps*

The time-dependent dynamics extracted from 2D IR further confirms that the spectral signatures at early times are dominated by distinct polariton nonlinearities, different from the spectra at 25 ps, which are mostly due to the population of molecular dark modes. The most notable difference between the early and late time dynamics is that at early times, the pump-probe and 2D IR spectral features are absorptive and the spectra oscillate as a function of $t_2$. These oscillations are revealed in the $t_2$ dynamics of the 2D IR spectral response when the pump is resonant with polariton frequencies, i.e., when $\omega_1=\omega_{UP}$ (Fig. 4a) or $\omega_1=\omega_{LP}$ (see section S2.4). The oscillation period is c.a. 0.8 ps (41.7 cm$^{-1}$ in frequency domain), which agrees well with the independently measured Rabi splitting of 38 cm$^{-1}$. Therefore, we assign these periodic dynamics to Rabi oscillations, which may be understood to arise from the evolution of polariton coherences $|UP\rangle\langle LP|$ and $|LP\rangle\langle UP|$, or equivalently from the energy exchange between the cavity electromagnetic field and the collective vibrational polarization.

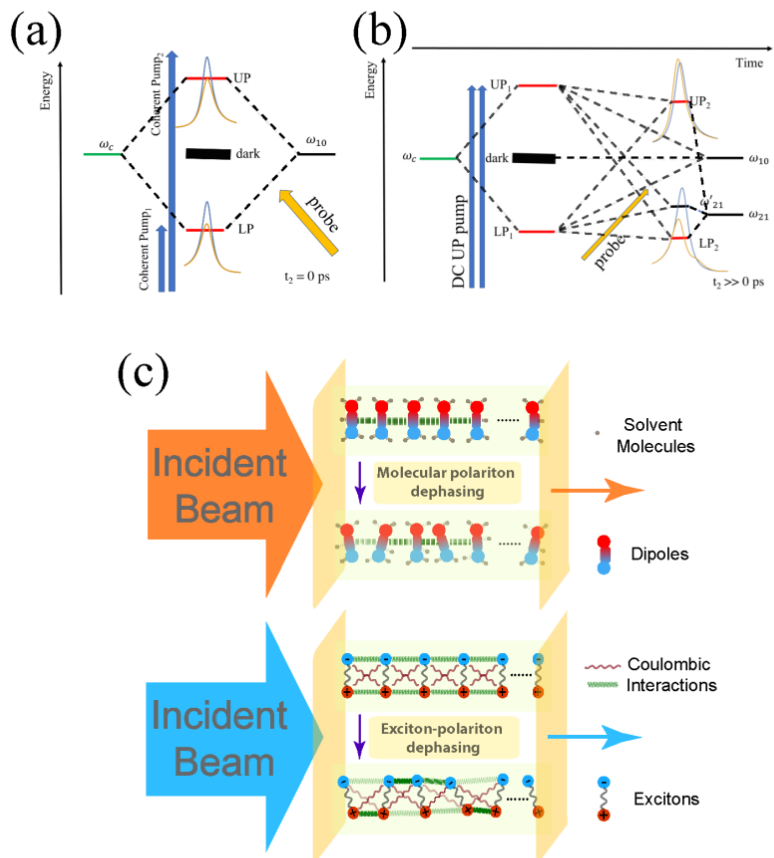

*Figure 5. Graphical representation of coherent (AC) and incoherent (DC) optical response (a) Energy level diagram showing the processes giving rise to the polariton bleach and AC signal in Fig. 4, where pumping generates LP-UP coherences and thus Rabi oscillations, and polariton bleach is observed when the probe interacts with the system at a sufficiently short pump-probe waiting time ($t_2 < 5$ ps) after pumping. Blue lineshapes represent linear response signals, while yellow show the reduced transmission characteristic of the polariton bleach effect. (b) Scheme showing the DC or long-time ($t_2 \gg 3$ ps) pump-probe response, where polariton population is generated and subsequent decay into the dark reservoir leads to changes in the collective oscillator strength due to a reduction in the average number of molecules in the ground-state. Blue lineshapes represent linear response signals while yellow give the pump-probe polariton lineshapes. (c) Schematic representation of nonlinear dephasing mechanisms for dipole-active vibrational (top) and inorganic exciton-polaritons (bottom). Strong electron-hole Coulomb interactions induce fast dephasing of the Wannier-Mott excitons of inorganic exciton-polaritons. These forces tend to be much less relevant in the molecular case, since the excitations (dipoles) interact weakly. Therefore, nonlinear molecular vibrational dephasing arises from inter or intramolecular anharmonic interactions.*

By applying a Fourier filter at the Rabi frequency to a series of $t_2$-dependent 2D IR spectra (see Supplementary Materials Sec. S2.4) of the 25-μm system, we extract a signal component which oscillates as a function of $t_2$ (referred to as AC) and isolate it from the non-oscillatory (DC) part. In Fig. 4a, we provide the time-dependent evolution of the AC signal, and in Fig. 4b we show examples of AC and DC spectral cuts obtained from 2D IR spectra at $t_2 = 0.8$ ps. The AC features at the LP and UP frequencies (Fig. 4b) are absorptive and have almost equal intensity matching the lineshape of the polariton bleach (Fig. 3b). Conversely, the DC spectra show a dispersive shape at the UP and much stronger nonlinear response at the LP (Fig. 4b), agreeing with our previous model(*6*, *14*) for late-time pump-probe response (Fig.3c). Because the AC signal requires the evolution of a coherent superposition of LP and UP (or equivalently, of |UP⟩⟨LP| and |LP⟩⟨UP| coherences) during t$_2$, the corresponding nonlinear resonances can be ascribed to a small subset of Feynman diagrams including coherence evolution (see Supplementary Materials Sec. S3.2). Interestingly, this implies that signal pathways containing polariton population (represented by |LP⟩⟨LP| and |UP⟩⟨UP|) are *inessential* for the generation of polariton bleach. In summary, our experiments allow clear identification of two regimes of qualitatively distinct nonlinear polariton phenomena: at *early* times (< 5 ps, which is roughly the polariton dephasing time), a nearly-symmetric bleaching of transmission is observed with a lineshape essentially indistinguishable from the AC signal (Fig. 5a), while at *late* times, (> 5 $ps$) the time-dependent nonlinear signal is dominated by DC dynamics arising from the existence of ground state population bleach and excited dark-states (Fig. 5b).

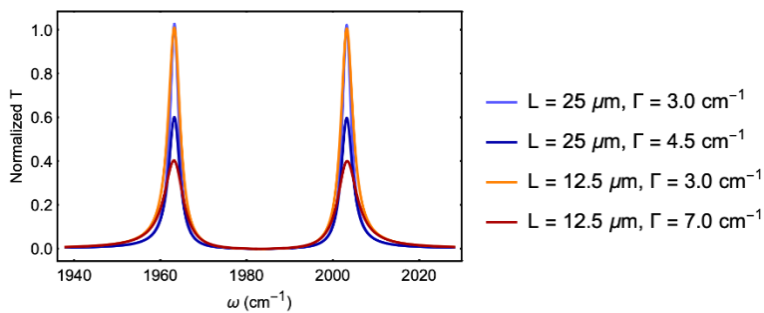

*Figure 6. Simulated polariton transmission from a classical phenomenological model where the effect of the pump is to change the molecular homogeneous dephasing rate, here represented by the FWHM, Γ. This process gives rise to enhanced polariton absorption and reduced transmission, reproducing the decrease in polariton transmission as observed in Fig. 1b and c.*

Given that the Rabi splitting is nearly invariant when the polariton bleach regime dominates the nonlinear response and only coherent excitations lead to AC signals, pump-induced oscillator strength loss can be neglected at early times, and the polariton bleach can be modeled by a phenomenological semiclassical theory(*32*) (see Supplementary Materials Sec. S3.3). In this description, the bleaching is reproduced when the molecular homogeneous broadening linewidth (full width at half maximum (FWHM), Γ) is changed from 3 cm$^{-1}$ to 4.5 cm$^{-1}$ for the 25-μm microcavitiy and from 3 cm$^{-1}$ to 7 cm$^{-1}$ for the 12-μm microcavity (Fig. 6). We note that the fitting only considers the strong coupling between fundamental mode (1983 cm$^{-1}$) because the coupling between 1 → 2 asymmetric stretch transitions of W(CO)$_6$ (1968 cm$^{-1}$) and the cavity mode is small due to large detuning and negligible population of reservoir modes at early times (*14*). This result, together with the insights gained from the 2D IR studies described above, is in sharp contrast to the traditional view of molecular pump-probe spectroscopy, where the nonlinear optical response is essentially modulated by excited-state population. Here, the essential mechanism is dephasing which becomes exacerbated to generate broadened polariton linewidths and reduced probe transmission when the pump is resonant with polaritons.

The polariton bleach phenomenon may also be visualized via a quantum mechanical analysis using double-sided Feynman diagrams (*22*), which we provide in detail in the Supplementary Materials Sec. S3.2. These diagrams represent the quantum pathways associated to the four peaks observed in the AC 2D IR spectra. Each AC peak has a stimulated emission (SE) and an excited-state absorption (ESA) pathway associated with the fundamental |g⟩ → |LP⟩ (or |g⟩ → |UP⟩ ) or combination band |LP⟩ → |UP, LP⟩ (or |UP⟩ → |UP, LP⟩) transitions. The phases corresponding to these pathways are opposite in sign. Therefore, if the fundamental and overtone transitions had equal oscillator strength, transition frequency, and linewidth (as is the case in harmonic systems), their amplitudes would interfere destructively leading to a vanishing 2D IR signal. In the nonlinear infrared spectroscopy of conventional molecular systems, the 2D response is generally dominated by features resulting from the anharmonic shift in frequency of the ESA which leads to incomplete cancellation of the stimulated emission (SE) and ground state bleach (GSB) amplitudes. This paradigm does not apply to the observed polariton bleach since no anharmonic energy shifts are visible in the recorded AC spectra; rather, the anharmonicity seems to be expressed in the linewidths, corresponding to nonlinear coupling to the environment. In particular, there is an incomplete cancellation of Feynman diagrams which is induced by a nonlinear dephasing mechanism that affects the $|LP, UP\rangle\langle LP|$ and $|UP, LP\rangle\langle UP|$ coherences but does not act on $|UP\rangle\langle g|$ or $|LP\rangle\langle g|$.

## DISCUSSION.

The observed nonlinear molecular polariton dephasing provides yet another instance where physical processes become relevant when light-matter interactions transition from the weak to the strong coupling regime. From a *microscopic* perspective, the polariton bleach arises from nonlinear dephasing activated by strong polariton excitation (~10 mJ/cm2). This effect has been observed before in inorganic semiconductor exciton-polaritons under intense pumping (*23–25*). In these systems, the rapid dephasing is induced by Coulomb scattering (Fig. 5c), which leads to enhanced broadening of the exciton homogeneous linewidth also detected in bare exciton pump-probe spectroscopy (*32*). For most chemical systems, the intermolecular Coulomb interactions can be well described by nonradiative dipole-dipole forces which decay with the inverse sixth-power of the distance. These interactions are likely too weak for the vibrational modes here studied due to large average distance between $W(CO)_6$ molecules (roughly 3 nm when the concentration of $W(CO)_6$ is $\cong$ 40 mM) and small effective dipole moments ($\cong$ 1D for the CO asymmetric stretch of $W(CO)_6$) (*33*).

We are led to conclude that alternative microscopic mechanisms induce the nonlinear dephasing manifested in the molecular polariton pump-probe response (Fig. 5c). Given that the bleach effect is clearest in the AC signal (Figs. 4a and 4b), we can obtain insight into this phenomenon by analyzing the contributing Feynman diagrams in Supplementary Materials Sec. S3.2. This indicates that the AC response is particularly sensitive to the nonlinear dynamics of the *two-particle* state $|LP, UP\rangle$. This many-body state undergoes efficient irreversible decay due to its resonance with a macroscopic collection of states containing two vibrational quanta in the same or different molecules. The dephasing could be accelerated by the triple degeneracy of the CO asymmetric stretch which gives an enhanced phase space for relaxation potentially mediated by the local fluctuations of the weak liquid-phase solute-solvent forces (*34*). Nevertheless, any nonlinear intra or intermolecular interaction can contribute to the observed nonlinear dephasing by assisting the $|LP, UP\rangle$ decay.

The dependence of polariton optical nonlinearities on macroscopic properties (cavity longitudinal length and molecular concentration) is a result of the hybridization of a collection of molecular transitions with macroscopic cavity modes. The long polariton coherence length makes their optical nonlinearities particularly sensitive to macroscopic descriptors. At short pump-probe delay times, we consistently observed reduced polariton transmission due to nonlinear dephasing. As the cavity volume and the number of molecules in the polariton coherence volume decrease, polariton nonlinearities become stronger. This trend conforms with the notion that giant nonlinearities (e.g., photon blockade) emerge in the limit of single emitter strong coupling (unattainable in our experiment).

Multidimensional spectroscopy allowed us to tease out the mechanism that drives the oscillating polariton bleach and unambiguously show that it depends on polariton coherences. In contrast to exciton-polariton studies in the UV-visible range, where fluorescence measurements inform much of their excited state dynamics, 2D IR is a suitable method for the study of vibrational polaritons due to the *weak* fluorescence featured in this range of the electromagnetic spectrum.

The mid-IR nonlinear optical phenomena reported in this article rely on a delicate interplay between microscopic molecular anharmonicities and macroscopic electromagnetic parameters. The ease with which the described samples can be fabricated and handled at room-temperature, together with their unique effects that interpolate between the molecular and the photonic realms, make vibrational polaritons ideal platforms to design previously-undeveloped mid-IR nonlinear optical switches, control chemical reactivity, or serve as building blocks for quantum information processing applications. As an example, from our previous work(*13*, *14*), polariton nonlinear effects disappear within 200 ps. Thus, an all-optical modulator for mid-IR frequencies operating at repetition rate as high as 5 GHz can be developed based on polariton bleach. Our study has also evidenced that polariton nonlinearities can be straightforwardly tuned via macroscopic controls. This notion may prove particularly suitable in the context of microfluidic devices, where molecular concentrations, and therefore, optical nonlinearities can be changed on-the-fly. Alternatively, polariton nonlinear response could be actively modified by combining cavities with piezo, thermal, and optically responsive materials(*35*), thus providing additional features to integrated photonic circuitry. Finally, the polariton anharmonicities underlying the respective nonlinearities may have nontrivial implications to vibrational dynamics and thermally-activated chemical reactions (*36–41*). In summary, we envision the exotic nonlinear phenomena explored in this work will play an important role in the emergent field of molecular polaritonics in the upcoming years.

**MATERIALS AND METHODS.**

***Polariton Preparation.*** The $W(CO)_6$ (Sigma-Aldrich) /cavity system is prepared in IR spectral cell (Harrick) containing two dielectric $CaF_2$ mirrors separated by a 12 or 25 μm spacer and filled with $W(CO)_6$/hexane solution (concentration varies from 12 mM to 40 mM). The dielectric mirror has ~96% reflectivity. Because the Rabi splitting (20 to 38 $cm^{-1}$ for the investigated concentrations) is larger than the FHWM of both cavity (~11 $cm^{-1}$) and $W(CO)_6$ vibrational (~3 $cm^{-1}$) modes, the strong coupling criteria are satisfied.

***Pump Probe and 2D IR Spectroscopy.*** The spectrometer follows the pulse shaper enabled pump-probe 2D IR setup and a rotational stage is added to control beam incidence angle. In the 2D IR spectrometer, three IR pulses interact with a sample sequentially to create two vibrational coherences. The first coherence is characterized by scanning $t_1$. The second coherence introduces a macroscopic polarization which subsequently emits a third order IR signal, which is self-heterodyned and detected in frequency domain. Transient pump-probe signal could be obtained by simply setting $t_1 = 0$ fs. To obtain 2D IR spectra, the free-induction-decay (FID) in $t_1$ is numerically Fourier transformed.

**Acknowledgements:** We thank Dr. Jeffery C. Owrutsky for insightful discussions. We also thank Linfeng Chen for his help on designing Figure 1 and Jiaxi Wang for her help on the experiments. **Funding:** B.X., Y.L., W.X. are supported by AFOSR Young Investigator Program Award, FA9550-17-1-0094. B.X. thanks Roger Tsien Fellowship from UCSD Department of Chemistry and Biochemistry. The development of the theoretical model by R.F.R. is supported by AFOSR award FA9550-18-1-0289, while the analysis of the short time dynamics by R.F.R and J.Y.Z. was supported by NSF CAREER CHE 1654732. **Author Contributions:** W.X. supervised the overall research and J.Y-Z. supervised the theoretical work. B.X. and W.X. designed the experiments. B.X., Y.L. conducted the experimental work. B.X. and W.X. analyzed experimental data. A.D.D., B.B.S. provided cavity optics. R.F.R. and J.Y-Z developed theoretical work. B.X., R.F.R., J.Y-Z. and W. X. interpreted the experimental and theory results. B.X., R.F.R., A.D.D., B.B.S., J.Y-Z. and W. X discussed the results and contributed to the final manuscript. **Competing interests:** The authors declare that they have no competing interests. **Data availability.** All data needed to evaluate the conclusions in the paper are present in the paper and/or Supplementary Materials. Additional data related to this paper and other finding of this study are available from the corresponding author upon reasonable request.

*Corresponding author*

*Correspondence to Wei Xiong.

**Supplementary Materials.**

Supplementary notes and figures can be found in the supplementary Materials.

Supplementary Materials

# Part I. Experimental Methods

# Part II. Supporting Results

# Part III. Theory

# Part I. Experimental Methods

## S1.1 Sample Preparation

The W(CO)$_6$ (Sigma-Aldrich) /cavity system is prepared in an IR spectral cell (Harrick) containing two dielectric CaF$_2$ mirrors separated by a 5, 12 or 25-μm Teflon spacer and filled with W(CO)$_6$/hexane solution (concentration varies from 5 mM to 50 mM). The dielectric mirror has a ~96% reflectivity. Because the Rabi splitting (20 to 37 cm$^{-1}$) is larger than the full-width-at-half-max of both cavity (~11 cm$^{-1}$) and W(CO)$_6$ vibrational (~3 cm$^{-1}$) modes, the strong coupling criteria are satisfied.

## S1.2 Two-Dimensional Infrared Spectrometer

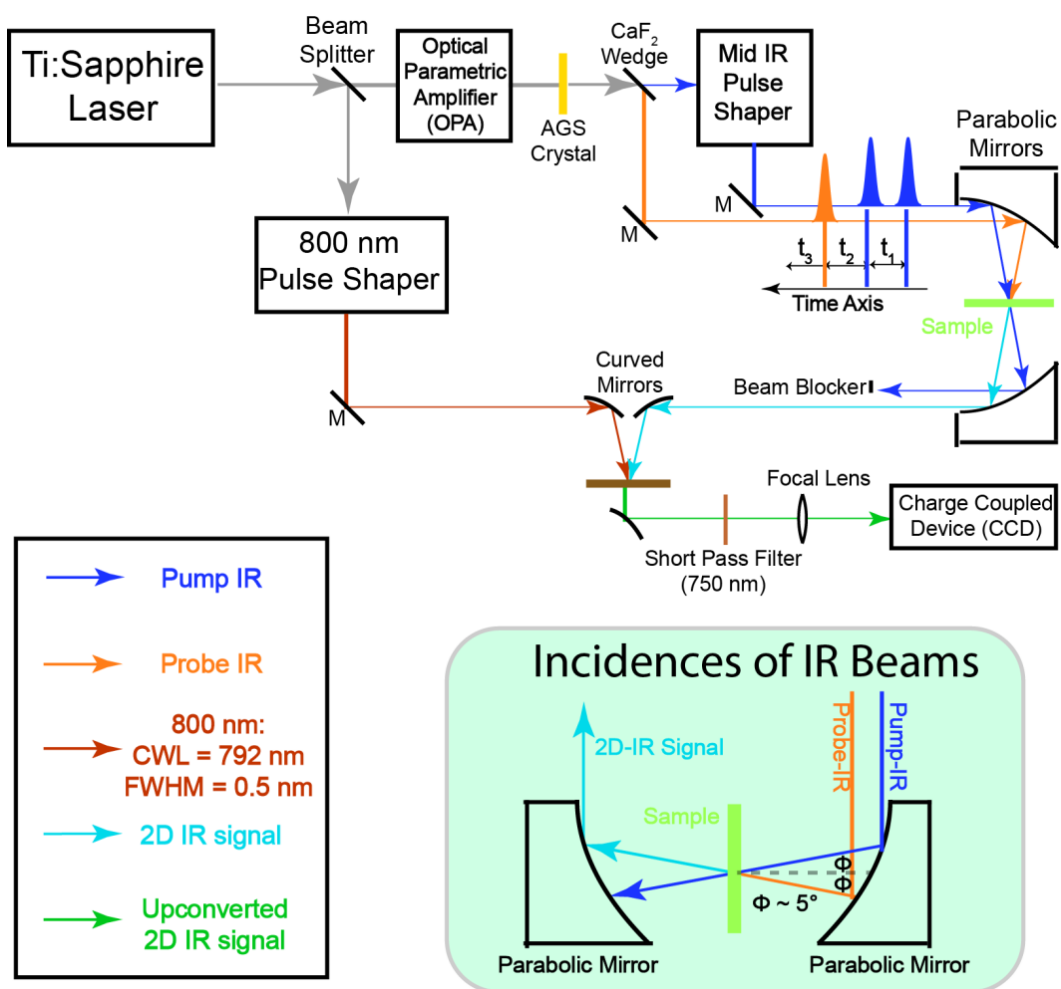

*Figure S1.* Scheme of two-dimensional infrared experimental setup, where the inset shows the incidence of pump and probe IR beams.

Two-dimensional infrared (2D IR) spectroscopy(*14*) is applied to investigate the light-matter interaction of a $W(CO)_6$/microcavity system. The setup scheme is shown in Fig. S1. 800-nm laser pulses (~35 fs, ~5 W, 1 kHz) generated by an ultrafast Ti:Sapphire regenerative amplifier (Astrella, Coherent) are sent into an optical parametric amplifier (OPA) (TOPAS, LightConversion) which outputs tunable near-IR pulses. The near-IR pulses are converted to mid-IR pulses through a difference frequency generation (DFG) process by a type II $AgGaS_2$ crystal (Eksma). After DFG, a $CaF_2$ wedge splits the mid-IR pulse into two parts: the 95% transmitted part is sent into a Ge-Acoustic Optical Modulator based mid IR pulse shaper (QuickShape, PhaseTech)(*42*) and is shaped to double pulses, which forms the pump beam arm; the 5% reflected is the probe beam. Both pump (~ 1.1 µJ) and probe (~ 0.2 µJ) are focused by a parabolic mirror (f = 10 cm) and overlap spatially at the sample. The output signal is collimated by another parabolic mirror (f = 10 cm) at a symmetric position and is upconverted to an 800-nm beam at a 5%Mg: $LiNbO_3$ crystal. The 800-nm beam that comes out of the OPA passes through an 800-nm pulse shaper which narrows its spectrum in the frequency domain (center wavelength of 791 nm and a FWHM of 0.5 nm or 9.5 $cm^{-1}$).

The pulse sequence is shown in Fig. S1. Two pump pulses and a probe pulse (pulse duration of 100~150 fs) interact with samples at delayed times ($t_1$, $t_2$ and $t_3$). After the first IR pulse, a vibrational coherence is generated, which is converted into a subsequent state by the second IR pulse and is characterized by scanning $t_1$ (0 to 6000 fs with 20 fs steps) using the mid IR pulse shaper. A rotating frame at $f_0$ = 1583 $cm^{-1}$ is applied to shift the oscillation period to 80 fs and to make the scanning step meet the Nyquist frequency requirement. After waiting for $t_2$, the third IR pulse (probe) is impinged on the sample, and the resulting macroscopic polarization emits an IR signal. This IR signal is upconverted by a narrow-band 800 nm beam. The upconversion process covers the $t_3$ time delay and the 800-nm pulse duration (full width at half maximum = 0.5 nm) determines the scanning length of $t_3$. The monochromator and CCD (Andor) experimentally Fourier transform the upconverted signal, thus generating a spectrum along the $\omega_3$ axis. Numerical Fourier transform of the signal along the $t_1$ axis is required to obtain the spectrum along $\omega_1$. The resulting 2D IR spectra are plotted against $\omega_1$ and $\omega_3$. The $t_2$ time delay is scanned by a computerized delay stage which is controlled by home-written LabVIEW programs to characterize the dynamic features of the system. A rotational stage is mounted on the sample stage to choose the tilt angle and, therefore, the wavevector of the driven polaritons. One special requirement for this experiment is that the rotation axis of the stage needs to be parallel to the incidence plane formed by the pump and probe beams. In this way, we ensure that the in-plane wavevectors, $k_{||}$, of pump and probe pulses are the same. The particular $k_{||}$ value the of pump and probe beams are determined by checking the 1D transmission polariton spectra of the pump and probe pulses before and after 2D IR acquisitions.

### S1.3 Three-Dimensional Fourier Transformation

2D IR spectra at early times ($t_2$ = 0 ~ 6 ps) show unambiguous oscillating features at the UP (see Fig. 4a) and LP frequencies (see Fig. S7a) and their period (0.8 ps) suggests these are Rabi oscillations (~38 $cm^{-1}$). This information provides key evidence of the coherence between the LP and the UP. To obtain the oscillatory part of the total 2D signal, we applied the 3D Fourier transform to the combined 2D matrices at different time delays. An additional frequency axis ($\omega_2$) is generated from applying FFT to the $t_2$ axis. Figures S2a and S2b contains the projections of the 3D FFT of the nonlinear signal into the $\omega_1$-$\omega_2$ and $\omega_3$-$\omega_2$, respectively. These figures clearly show features at $|\omega_2|$ = 38 $cm^{-1}$ (referred to as AC part), in addition to the non-oscillating part at $\omega_2$ ~ 0 $cm^{-1}$ (referred to as DC part). The DC and AC parts can be disentangled by applying a frequency filter to the 3D matrix followed by an inverse Fourier transform of the DC and AC parts. A representative result is given in Fig. 4b of the main text.

To further analyze the coherent dynamics, amplitudes of the AC part at $\omega_1 = \omega_{UP}/\omega_3 = \omega_{UP}$ and $\omega_1 = \omega_{LP}/\omega_3 = \omega_{UP}$ were extracted and plotted versus $t_2$ (Figs. S2c and S2d). The UP-UP oscillating trace exhibits nearly perfect dephasing dynamics and can be fitted with a single exponential giving a lifetime of ~2 ps. Conversely, the LP-UP trace deviates from pure dephasing dynamics, possibly because the overtone transitions of dark reservoir modes ($\upsilon_{12} \sim 1968$ cm-1 and $\upsilon_{23} \sim 1950$ cm-1) perturb the LP state.

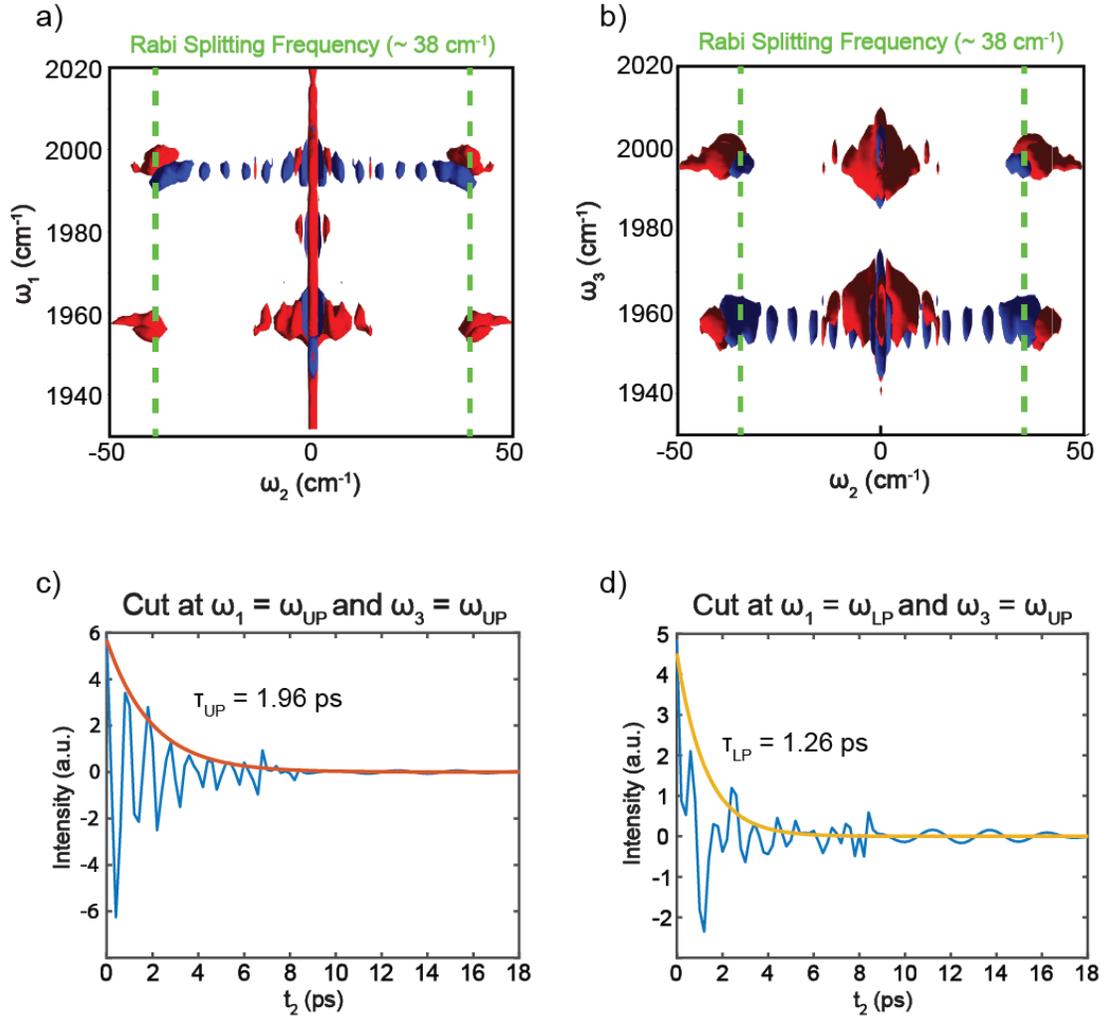

*Figure S2.* 3D FFT frequency domain ($\omega_1$-$\omega_2$-$\omega_3$) spectra: (a) View from the $\omega_1$-$\omega_2$ plane; (b) View from the $\omega_3$-$\omega_2$ plane; Dephasing traces of (c) cut at $\omega_1 = \omega_{UP}$ and $\omega_3 = \omega_{UP}$, and (d) cut at $\omega_1 = \omega_{LP}$ and $\omega_3 = \omega_{UP}$.

## S1.4 Nonlinear Signal Pump and Probe Power-dependence

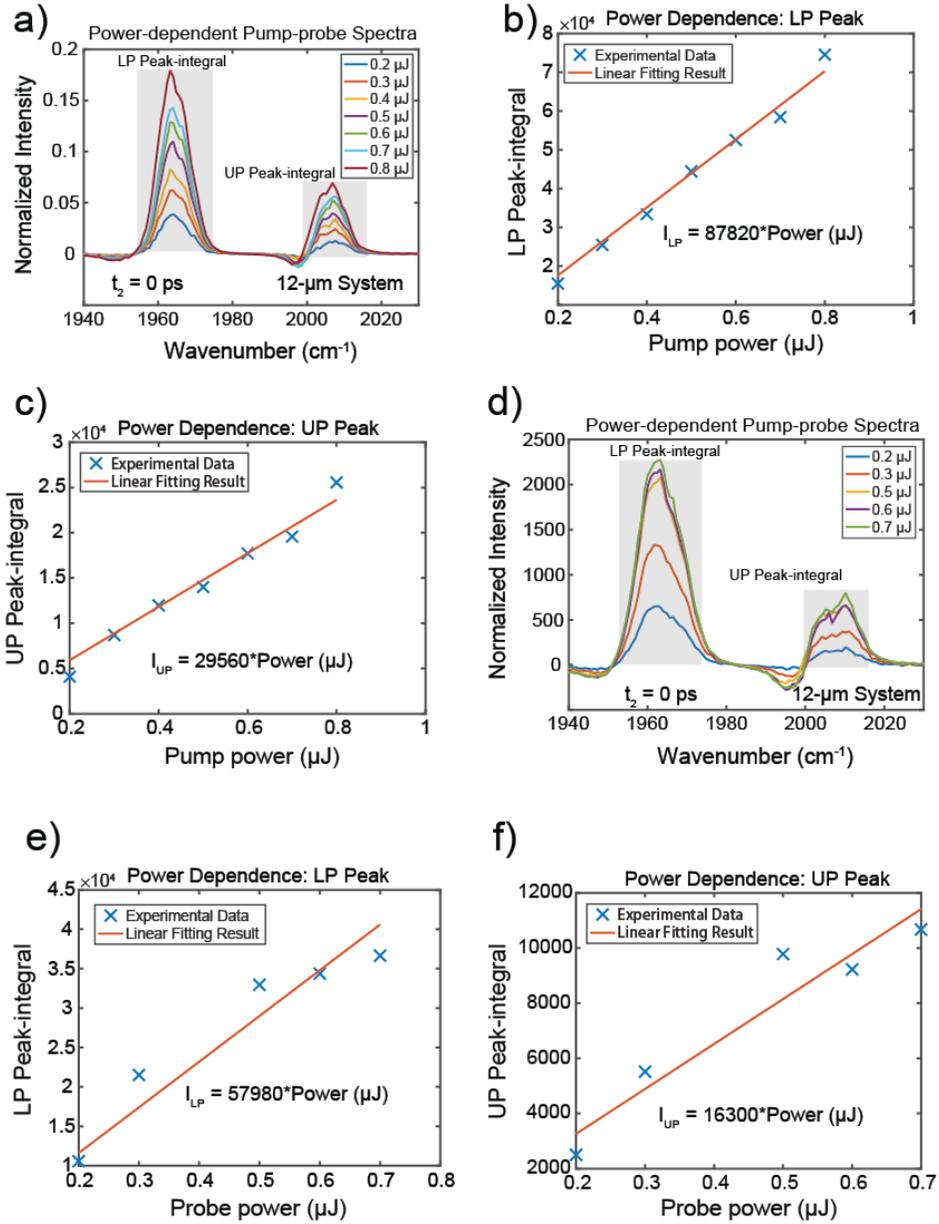

*Figure S3.* Pump power-dependence: (a) Pump-probe spectra at $t_2$ = 0 ps for 12-μm polariton system; (b) LP peak integrated signal; (c) UP peak integrated signal. Probe power-dependence: (d) Pump-probe spectra at $t_2$ = 0 ps for 12-μm polariton system; (e) LP peak integrated signal; (f) UP peak integrated signal. Both LP and UP peak integrated signals are proportional to pump/probe power, supported by the fitting equation shown as the insets.

To figure out the scaling of the 'polariton bleach' signal with respect to the pump intensity, we performed a series of power-dependent pump-probe experiments. The self-heterodyned third order signal (ΔT) can be expressed as the equations as follow

$$\Delta T = \left|Sig^{(3)} + Sig^{(1)}\right|^2 - \left|Sig^{(1)}\right|^2 = 2Re\left[Sig^{(3)} * \overline{(Sig^{(1)})}\right] + \left|Sig^{(3)}\right|^2,$$

$$\text{where } Sig^{(3)} \sim E^2_{pump} \cdot E_{probe} \text{ and } Sig^{(1)} \sim E_{probe}.$$

$$\Rightarrow \Delta T \propto E^2_{pump} E^2_{probe} + O(E^4_{pump} E^2_{probe})$$

Since $(E_{pump})_2$ and $(E_{probe})_2$ are proportional to the pump and probe power, respectively, the signal ($\Delta T$) is proportional to the power of both IR beams whenever higher-order response can be neglected. From the pump-power-dependent results (Fig. S3), it is clear that both LP and UP peak integrated signal intensities decrease as pump or probe power reduces. The integrated LP and UP peak intensities are roughly proportional to the pump or probe power (Fig. S6b, S6c, S6e and S6f). The linear relations between LP and UP peak-integrals and IR power have been shown as the insets of the corresponding sub-figures.

## Part II. Supporting Results

### S2.1 Transmission and Transient Pump-probe Spectra of Uncoupled Systems

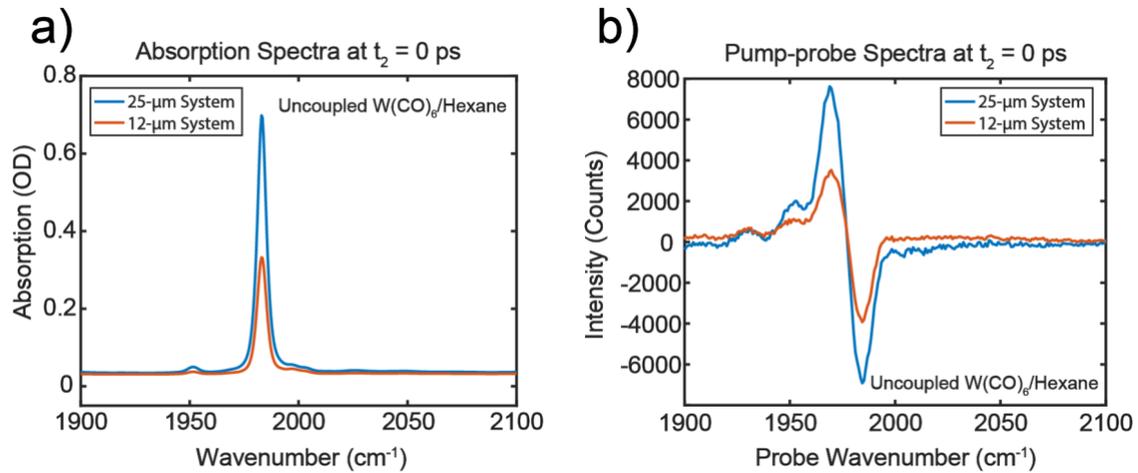

***Figure S4.*** *Results of control experiments for uncoupled W(CO)$_6$/hexane systems with both 12 and 25-micron cell longitudinal lengths at similar concentration: (a) Linear transmission spectra; (b) Pump-probe spectra.*

We performed the control experiments with uncoupled W(CO)$_6$/hexane systems with 12 and 25 μm spacers under similar conditions (concentration, IR power, etc). The transmission in Fig. S4a shows agreement with Beer's law, the absorption is doubled when the cavity longitudinal length is switched from 12-μm to 25-μm and the ratio between the absorption of 12-μm and 25-μm uncoupled systems is 2.03. In Fig. S4b, the pump-probe signal of 25-μm is roughly twice as large as the 12-μm pump-probe signal. Based on the fundamental peak (the only which is negative, around 1983 cm$_{-1}$), the ratio between 25-micron and 12-μm peak intensity is about 2.16, while the 1→2 overtone peak (at roughly probe wavenumber of 1968 cm$_{-1}$) intensity ratio is 1.82. Both ratios are close to 2. Therefore, this result suggests that the 3rd order signal (pump probe) is not enhanced relative to the linear signal (absorption) when the sample length is reduced by a factor of 2.

## S2.2 2D IR and Transient Pump-Probe Spectra and 2D-IR Spectral Cuts

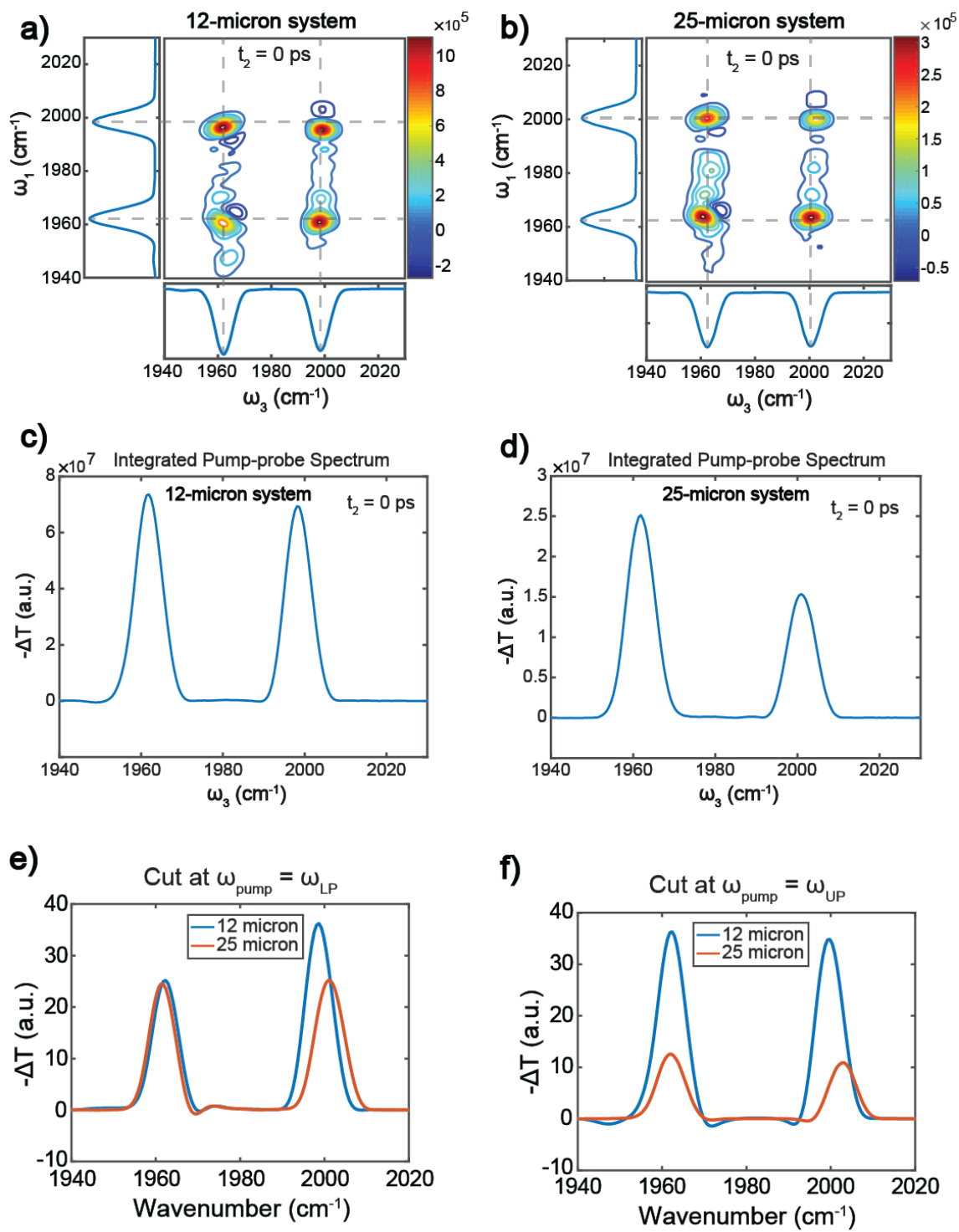

*Figure S5.* 2D IR spectra of the (a)12 μm and (b) 25 μm systems; Integrated pump-probe spectra of the 12 μm (c) and (d) the 25 μm systems. Note that the color bars of (a) and (b) indicate that 12-micron system has significantly stronger nonlinearities than 25-micron system. 2D IR Spectral cuts of the 12 and 25 μm systems for comparison at (a) $\omega_{pump} = \omega_{LP}$; (b) $\omega_{pump} = \omega_{UP}$. $t_2 = 0$ ps

**S2.3 2D IR Spectra for Various Molecular Concentrations**

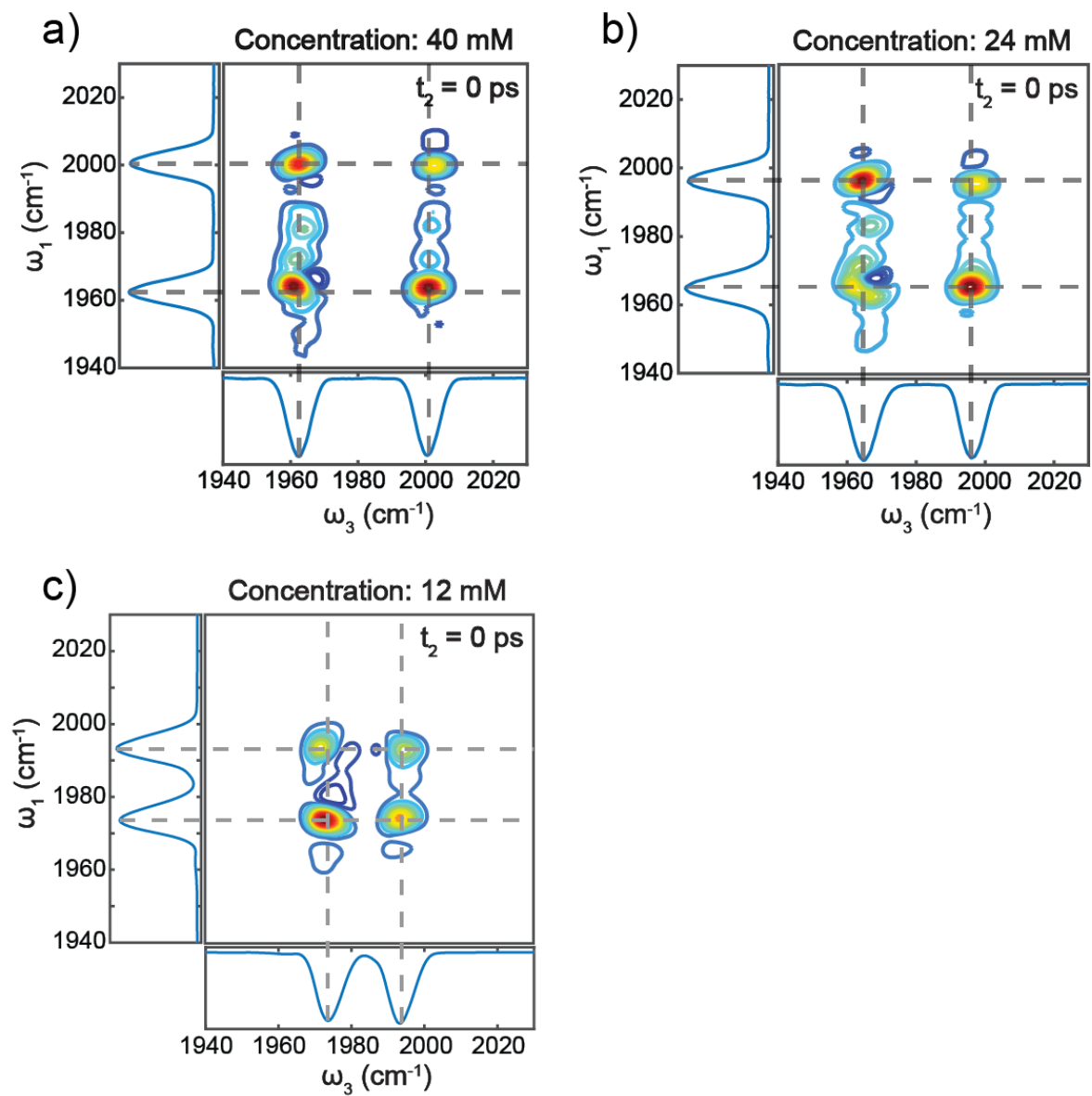

***Figure S6.*** *2D IR spectra of the 25 μm system at $t_2 = 0$ ps with various concentrations: (a) 40 mM; (b) 24 mM;*

## S2.4 Early-time Dynamics of 2D IR LP and Dark Mode Spectral Cuts

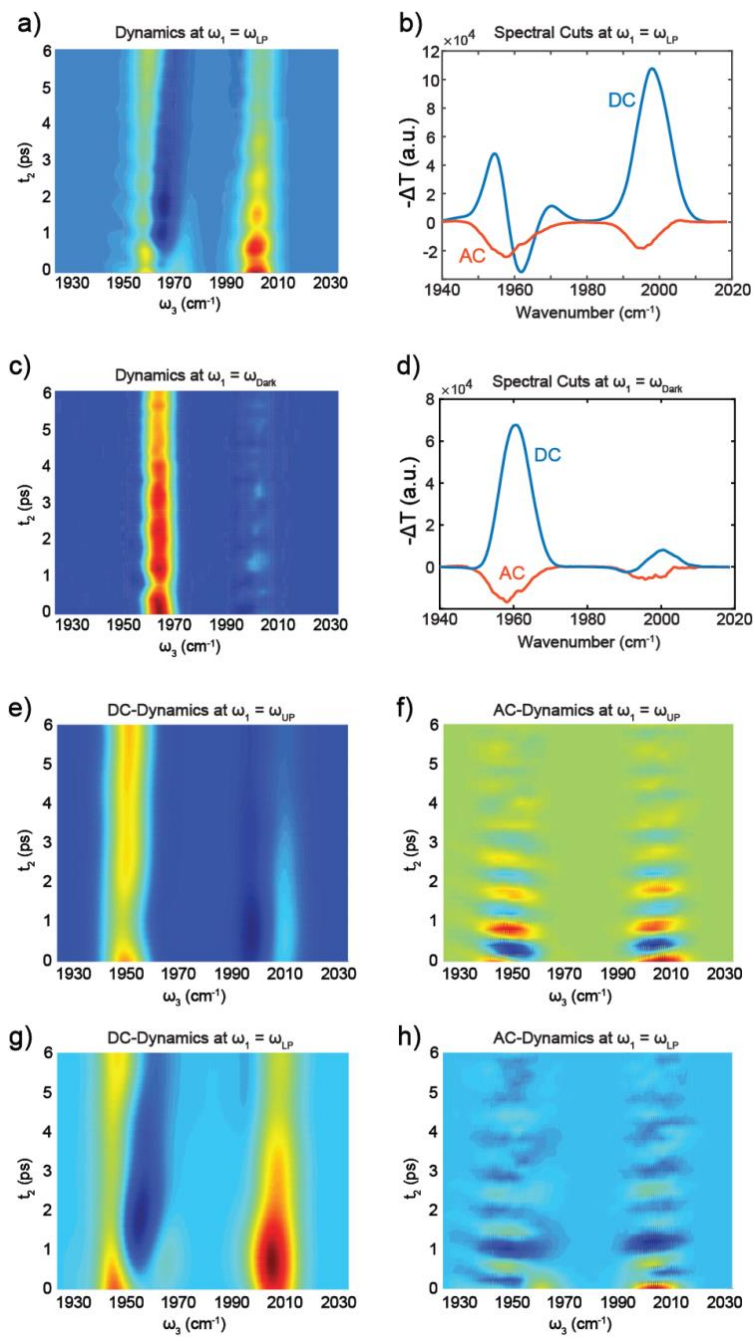

*Figure S7.* Early-time dynamics of 2D IR spectral cuts at (a) $\omega_1=\omega_{LP}$ and (c) $\omega_1=\omega_{Dark}$ between -2 to 7 ps. AC and DC differential transmission spectral cuts at (b) $\omega_1=\omega_{LP}$, and (d) $\omega_1=\omega_{Dark}$ at $t_2=0.8$ ps. Dynamics of (e) DC-component at $\omega_1=\omega_{UP}$, (f) AC-component at $\omega_1=\omega_{UP}$, (g) DC-component at $\omega_1=\omega_{LP}$, and (h) AC-component at $\omega_1=\omega_{LP}$.

Figures S7a and S7c show the waiting-time ($t_2$) dynamics of spectral cuts at $\omega_1=\omega_{LP}$ and $\omega_1=\omega_{Dark}$, respectively. We also employed the 3D-FFT (Sec. S1.3) to extract the AC and DC components of each of the mentioned spectral cuts. In Figs. S7b and S7d, we show the AC and DC signal contributions to nonlinear signal obtained with $\omega_1=\omega_{LP}$ and $\omega_1=\omega_{Dark}$ at $t_2 = 0.8$ ps, respectively.

Compared to the dynamics of the UP-cut (Fig. 4a), the waiting-time dependent signal measured at the LP-cut (Fig. S7a) has significantly stronger integrated DC component, although the oscillatory bleach feature is still obvious (Fig. S7b). The relative phase difference between the LP-AC (Fig. S7a) and the UP-AC (Fig. 4b) signals is due to the heterodyne detection system which allows us to extract the relative phase between the signals obtained with LP and UP pumping. The enhanced integrated DC signal for an LP-resonant pump is likely a result of the perturbation of the LP by overtone transitions of $W(CO)_6$ dark modes, especially that with $\omega_{12} \sim 1968$ cm$^{-1}$, which is nearly resonant with the LP ($\omega_{LP} = 1964$ cm$^{-1}$), and thus perturbs the polariton response even at early-times. When the pump frequency equals that of the bare molecule (dark mode), the oscillatory part of the nonlinear signal is almost negligible (Fig. S7c). After Fourier filtering, it does show a very weak AC component (Fig. S7d), which may be a result of spectral overlap with the oscillating polariton spectral features. As discussed in the main text and Sec. S3.2, $|UP\rangle\langle LP|$ and $|LP\rangle\langle UP|$ coherence states are required for the observation of Rabi oscillations (AC part). Due to spectral congestion and overlaps, the tail of the AC component of LP and UP peaks can also induce dark mode spectral features to "oscillate" weakly. Nevertheless, as clearly shown in the 3D IR plot (Figs. S2a and S2b), the Rabi oscillation peaks near $\omega_2=38$ cm$^{-1}$ only show up at LP and UP transitions, but not at dark mode resonances. This provides strong evidence that dark modes are not involved in Rabi oscillation. In summary, the weakly oscillatory features in the dark mode dynamics, and the stronger contamination of the LP response by dark mode transitions indicate the UP spectral cuts are the most suitable for the analysis of pure vibrational polaritonic response, as performed in the main text.

The early-time dynamics of AC and DC parts of both LP and UP spectral cuts are shown in Fig. S7e, f, g and h. The DC parts (Fig. S7e and g) of both LP and UP cuts show the expected peak-shifts which evolve continuously. The AC parts (Fig. S7f and h), on the other hand, are mostly composed of a convolution of oscillations with dephasing traces (see SI section 1.3 Three-dimensional Fourier Transform for more details). By comparing the DC and AC dynamics, it is clear that the oscillating part which contributes significantly to the 'polariton bleach' signal would dephase within approximately 3 ps, matching with the cavity lifetime. Conversely, the DC parts continue to evolve even after 3 ps. Those dynamic traces further support our statements.

## S2.5 Two-Component Spectral Fitting of Absorptive Pump-Probe Spectra

In this section, we show that the measured nonlinear signals obtained from the LP and UP (25 μm system) 2D IR spectral cuts at zero waiting-time cannot be reproduced with a simple spectral fitting based on previous studies(*6*, *13*, *14*).

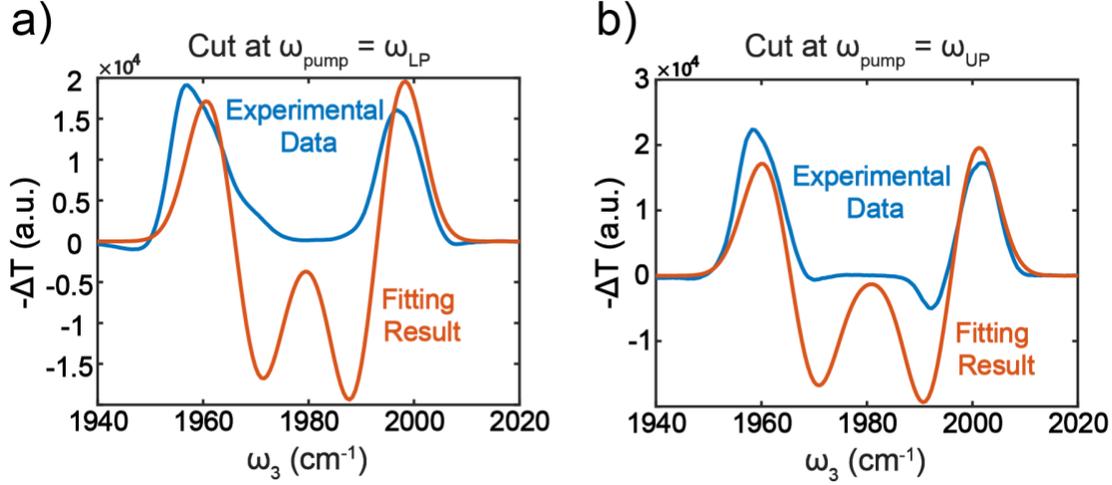

***Figure S8.*** *Spectral fitting of 25-μm systems with $t_2 = 0$ ps at (a) $\omega_1=\omega_{LP}$ and (b) $\omega_1=\omega_{UP}$. The fitting model is composed of an absorptive feature on the LP (due to the interference of the LP with dark overtone transitions) and the derivative lineshape at the LP/UP frequencies (due to Rabi splitting contraction). The results indicate that the combined effect of dark mode overtone absorption and Rabi splitting contraction is insufficient to accurately model the experimental features at early times.*

A two-component spectral fitting was applied to the LP/UP spectral cuts of 2D-IR spectrum of 25-μm system at $t_2 = 0$ ps, using the following equation

$$Spec_{fit}(x) = \underbrace{\left\{\alpha_1\left[a_0 e^{-\left(\frac{x-b_0}{c_0}\right)^2}\right]\right\}}_{(1)} + \underbrace{\left\{\alpha_2 \sum_{i=1}^{2}\left[a_i e^{-\left(\frac{x-b_i}{c_i}\right)^2} - a_i e^{-\left(\frac{x-b_i+\Delta x}{c_i}\right)^2}\right]\right\}}_{(2)} \quad (1)$$

In this equation, terms (1) and (2) represent spectral components corresponding to dark mode overtone response and the Rabi splitting contraction, respectively. The probed frequency is given by $x$, and $a$, $b$ and $c$ are the amplitude, resonance frequency and width of a Gaussian peak; $\Delta x$ is the amount of peak-shift and $\alpha_1$ and $\alpha_2$ are the coefficients which modulate the compositions of the two components.

**Table S1.** Fitting Parameters

| States of Spectral Cuts | Parameters for Gaussian Peaks | | | | | | Compositions | | Polariton Peak-shift (cm-1) |
|---|---|---|---|---|---|---|---|---|---|
| | Amplitudes | | Resonance Frequencies (cm-1) | | Widths (cm-1) | | | | |
| | $a_1$ | $a_2$ | $b_1$ | $b_2$ | $c_1$ | $c_2$ | $\alpha_1$ | $\alpha_2$ | $\Delta x$ |
| UP | 17550 | 20000 | 1960.5 | 2001 | 8.1 | 7.7 | 0.1 | 0.5 | 10 |
| LP | 17550 | 20000 | 1960.5 | 1998 | 8.1 | 7.7 | 0.1 | 1 | 10 |

From Figs. S8a and S8b, it is clear the spectral cuts from the experimental data cannot be fitted with only the two components mentioned in equation (6). Hence, an additional component that represents the polariton-bleach behavior would be needed.

## Part III. Theory

### S3.1 Scaling of Polariton Nonlinearities

In this section, we show that when the effects of dark modes can be disregarded, the dominant coherent polariton nonlinear interactions induced by quartic couplings (the most relevant interactions at low energies for homogeneous and isotropic systems like the molecular solution studied here) leads to inverse scaling with respect to the cavity longitudinal length *due to polariton delocalization*. We first demonstrate explicitly the inverse dependence with cavity longitudinal length and molecular concentration of vibrational polariton nonlinearities induced by local mechanical and electrical molecular anharmonicity. The discussion is subsequently generalized to include more general types of molecular nonlinearities, including the nonlinear dephasing mechanism which we propose as the source of the polariton bleach effect. The discussion in this section is limited to polariton-polariton interactions relevant at short times relative to their lifetime

Let a planar infrared cavity consisting of two ideal parallel metal mirrors with surface area $S$ separated by longitudinal length $L$ host $N$ molecules homogeneously distributed across the volume $V = SL$. The TE and TM modes of the cavity are obtained by using periodic boundary conditions for the electromagnetic field along the $x$ and $y$ directions and metal-dielectric interface boundary conditions from Maxwell equations. For the sake of simplicity, we retain in what follows only a single TE cavity photon band with modes parametrized by $\mathbf{k} = (k_x, k_y)$ having resonance frequency $\omega(\mathbf{k})$(*43*). The vibrational frequency and position of the molecule $i$ in the set $\{1,...,N\}$ are given by $\omega_i$ and $\mathbf{r}_i = (x_i, y_i, z_i)$, respectively. The light-matter interaction is treated within the dipole approximation in the rotating-wave-approximation (only interactions which conserve the excitation-number are included)(*43*). In the *absence* of anharmonicity, the Hamiltonian of the hybrid system is given by:

$$H_0 = \sum_{i=1}^{N} \hbar\omega_i a_i^\dagger a_i + \sum_{\mathbf{k}} \hbar\omega(\mathbf{k}) b^\dagger(\mathbf{k}) b(\mathbf{k}) + \sum_{i=1}^{N} \sum_{\mathbf{k}} \left[ \bar{g}_{i\mathbf{k}} a_i^\dagger b(\mathbf{k}) + g_{i\mathbf{k}} b^\dagger(\mathbf{k}) a_i \right] \quad (2)$$

where $a_i$ and $b(\mathbf{k})$ are the molecular and photonic annihilation operators, $g_{i\mathbf{k}}$ is the coupling constant for the interaction between molecule $i$ and cavity mode $\mathbf{k}$ and $\overline{g_{i\mathbf{k}}}$ is its complex conjugate. The detailed form of $g_{i\mathbf{k}}$ can be found elsewhere(*44*). For our purposes, all that is relevant is that

$$g_{i\mathbf{k}} \propto e^{-i\mathbf{k}\cdot\mathbf{r}_i} \sqrt{\omega_\mathbf{k}} \quad (3)$$

where $\mathbf{k} \cdot \mathbf{r}_i = (k_x, k_y) \cdot (x_i, y_i)$. Disregarding the inhomogeneous broadening of the molecular vibrations (i.e., supposing all molecular frequencies are equal to $\omega_0$), the polariton modes of this Hamiltonian can be obtained straightforwardly performing a canonical transformation of the local molecular operators into a collective basis adapted to the form of the light-matter interaction given above(*8*). Specifically, the following *bright* molecular operators are defined:

$$a(\mathbf{k}) = \sum_{i=1}^{N} \frac{g_{i\mathbf{k}}}{g_\mathbf{k}} a_i, \quad g_\mathbf{k} = \sqrt{\sum_{i=1}^{N} |g_{i\mathbf{k}}|^2} \propto \sqrt{N} \quad (4)$$

In this model, there exists also $N_d = N - N_k$ dark modes (where $N_k \propto S$ is the total number of cavity photon modes included in the effective description of the electromagnetic modes) which form the complement to the bright operators $a(\mathbf{k})$ in the space of annihilation operators(8). In the collective basis, the quadratic part of the Hamiltonian is written as:

$$H_0 = \hbar\omega_0 \sum_{\mathbf{k}} a^\dagger(\mathbf{k})a(\mathbf{k}) + \sum_{\mathbf{k}} \hbar\omega(\mathbf{k})b^\dagger(\mathbf{k})b(\mathbf{k}) + \sum_{\mathbf{k}} \left[\bar{g}_{\mathbf{k}} a^\dagger(\mathbf{k})b(\mathbf{k}) + g_{\mathbf{k}} b^\dagger(\mathbf{k})a(\mathbf{k})\right] + \hbar\omega_0 \sum_{d=1}^{N_d} \alpha_d^\dagger \alpha_d \quad (5)$$

where the operators $\alpha_d$ are the annihilation operators of dark modes. The bright normal modes of the above Hamiltonian are given by the lower (LP) and upper polaritons (UP) with annihilation operators written as(8):

$$\alpha_{UP}(\mathbf{k}) = \cos(\theta_\mathbf{k}/2)a(\mathbf{k}) + \sin(\theta_\mathbf{k}/2)b(\mathbf{k}) \quad (6)$$

$$\alpha_{UP}(\mathbf{k}) = -\sin(\theta_\mathbf{k}/2)a(\mathbf{k}) + \cos(\theta_\mathbf{k}/2)b(\mathbf{k}) \quad (7)$$

where the specific form of the mixing angles $\theta_\mathbf{k}$ and polariton frequencies $\omega_{LP}(\mathbf{k})$ and $\omega_{UP}(\mathbf{k})$ are inessential for our discussion.

In order to describe polariton nonlinearities we introduce molecular anharmonicity to the hybrid cavity Hamiltonian. At low energies (i.e., when only the first few excited-states are probed) only cubic and quartic nonlinearities are relevant(22) Retaining only those interactions that preserve the excitation number operator (consisting of the sum of the photonic and phononic number operators) and disregarding nonlocal interactions or interactions with the environment, the nonlinear part of the Hamiltonian can be written generically as[7]:

$$H_I = \Delta \sum_{i=1}^{N} a_i^\dagger a_i^\dagger a_i a_i + \eta \sum_{i\mathbf{k}} \left[g_{i\mathbf{k}} a_i^\dagger a_i^\dagger a_i b(\mathbf{k}) + \bar{g}_{i\mathbf{k}} b^\dagger(\mathbf{k}) a_i^\dagger a_i a_i\right] \quad (8)$$

where $\Delta$ parametrizes the mechanical anharmonicity (deviation from harmonic energy spectrum of a single vibration), $\eta$ quantifies deviations of the $i$th molecule vibrational transition dipole function from linearity with respect to displacement of the molecular mode from its equilibrium position. In the hybrid cavity *normal-mode* basis, $H_I$ can be written as a sum of polariton-polariton, polariton-dark mode, and dark-dark interactions. The polariton-polariton interactions take the following form:

$$H_I^{\text{pol-pol}} = \sum_{p_1 p_2 p_3 p_4}^{\text{LP,UP}} \sum_{\mathbf{k}_1 \mathbf{k}_2 \mathbf{k}_3 \mathbf{k}_4} V_{p_1 p_2, p_3 p_4}^L(\mathbf{k}_1, \mathbf{k}_2, \mathbf{k}_3, \mathbf{k}_4) \alpha_{p_1}^\dagger(\mathbf{k}_1) \alpha_{p_2}^\dagger(\mathbf{k}_2) \alpha_{p_3}(\mathbf{k}_3) \alpha_{p_4}(\mathbf{k}_4)$$

(9)

where the $p_i$'s correspond to either LP or UP and $V_{p_1 p_2 p_3 p_4}^L(\mathbf{k_1}, \mathbf{k_2}, \mathbf{k_3}, \mathbf{k_4})$ are the coupling constants for the corresponding polariton-polariton interactions. The nonlinear coupling arising from mechanical anharmonicity satisfies the following relations:

$$V_{p_1 p_2, p_3 p_4}^\Delta(\mathbf{k}_1, \mathbf{k}_2, \mathbf{k}_3, \mathbf{k}_4) \propto \Delta \sum_{i=1}^{N} \frac{\bar{g}_{i\mathbf{k}_1}}{g_{\mathbf{k}_1}} \frac{\bar{g}_{i\mathbf{k}_2}}{g_{\mathbf{k}_2}} \frac{g_{i\mathbf{k}_3}}{g_{\mathbf{k}_3}} \frac{g_{i\mathbf{k}_4}}{g_{\mathbf{k}_4}}$$

$$\propto \Delta \sum_{i=1}^{N} \frac{e^{-i(\mathbf{k}_1 + \mathbf{k}_2 - \mathbf{k}_3 - \mathbf{k}_4) \cdot \mathbf{r}_i}}{N^2}$$

$$\approx c_\Delta \frac{\Delta}{N} \delta_{\mathbf{k}_1 + \mathbf{k}_2 - \mathbf{k}_3 - \mathbf{k}_4} \quad (10)$$

where $\delta_\mathbf{q}$ is the discrete delta function and $c_\Delta$ is a constant independent of particle number (in the thermodynamic limit). To obtain the last expression we used that the number of molecules is macroscopic, and these are distributed randomly inside the cavity so that the dominant contribution to the polariton-polariton interaction preserves the in-plane wave-vector (a similar result would have been obtained if we had assumed the wave-vector is not conserved, but only excitations with $k$ close to zero are driven by the pump). For the electrical anharmonicity term we find a similar result:

$$V^\eta_{p_1 p_2, p_3 p_4}(\mathbf{k}_1, \mathbf{k}_2, \mathbf{k}_3, \mathbf{k}_4) \approx \frac{\eta(\mathbf{k}_1 + \mathbf{k}_2 - \mathbf{k}_3)}{N} \delta_{\mathbf{k}_1 + \mathbf{k}_2 - \mathbf{k}_3 - \mathbf{k}_4} \quad (11)$$

where $\eta(\mathbf{k}_1 + \mathbf{k}_2 - \mathbf{k}_3)$ is a density-dependent coupling constant of the order of the Rabi splitting.

Further simplification arises from noting that the experimental nonlinear response is measured at $\mathbf{k} \approx 0$, and therefore the dominant contributions will have all $\mathbf{k}_i$ close to zero. Taking for simplicity only the (dominant) couplings between polaritons with the same wave-vector, the total nonlinear coupling can be written as as $\sum_k F(\mathbf{k})/N$, where $F(\mathbf{k})$ is independent of particle number. In the continuum limit of the photonic system, it follows that $\sum_k F(\mathbf{k})/N = (2\pi)^{-2} S \int d\mathbf{k}\, F(\mathbf{k})/N$. Given that $N = \rho S L$ where $\rho$ is the molecular density, we find that under the discussed assumptions the nonlinear polariton-polariton coupling constant is inversely proportional to $\rho L_z$. It follows that, for a fixed molecular density, increasing the cavity length, leads to an inversely proportional reduction in the dominant nonlinear polariton-polariton interactions. Similarly, for a fixed cavity longitudinal length, an increase in the molecular density reduces the magnitude of nonlinearities of the hybrid microcavity system.

While the results given above derive from the assumed nonlinearities in Eq. (8), they also apply to nonlinear interactions of the vibrational modes with other molecular degrees of freedom (bath modes). In the case of the W(CO)$_6$ solution, the bath contains, in addition to low-frequency intra and intermolecular modes, the carbonyl asymmetric stretch doublet corresponding to the vibrational modes orthogonal to that which interacts with the TE cavity photon. These modes are nearly-degenerate with the dark transitions, and significantly enhance the phase space for the decay of the two-body states $|LP(\mathbf{k}), UP(\mathbf{k})\rangle$. This process is favored by the fact that the energy of this state is close to twice the energy of the bare fundamental transition when $\mathbf{k}$ is such that the cavity photon is resonant with the molecular system. Thus, $|LP(\mathbf{k}), UP(\mathbf{k})\rangle$ is particularly sensitive to nonlinear dephasing mechanisms. The rate of decay of $|LP(\mathbf{k}), UP(\mathbf{k})\rangle$ into bath modes via resonant incoherent scattering is also proportional to $1/N$ as can be seen from the following argument. The perturbative rate of decay is proportional to $\sum_{i=1}^N \rho(F) \left|\left\langle F^{(i)} \middle| V_{SB}^{(i)} \middle| LP, UP \right\rangle\right|^2$, where the wave-vector dependence of the polariton states is implicit, $i$ labels each molecule, $V_{SB}^{(i)}$ denotes the interaction between the $i$th molecular vibration and the bath modes, and $F^{(i)}$ labels arbitrary bath states with density $\rho(F)$ which interact only with the $i$th molecule and that have approximately the same energy as the sum of the LP and UP energies. In the molecular basis, this term is proportional to $\sum_{i=1}^N \frac{\rho(F)}{N^2} \left|\left\langle F^{(i)} \middle| V_{SB}^{(i)} \middle| 2^{(i)} \right\rangle\right|^2 = \frac{\rho(F)}{N} \left|\left\langle F^{(m)} \middle| V_{SB}^{(m)} \middle| 2^{(m)} \right\rangle\right|^2$, where $m$ refers to an arbitrary molecule, and the last equality was obtained by assuming all molecules have equal matrix elements for their nonlinear interaction with the bath (mean-field approximation). It follows that nonlinear homogeneous dephasing induced by system-bath couplings also lead to the inverse scaling of nonlinearities with respect to the number of molecules, thus verifying the generic character of this polaritonic feature.

We conclude by highlighting that these considerations show that *the inverse scaling with 1/N of the vibrational polariton nonlinear response is fundamentally a consequence of the delocalization of the polariton across the molecular ensembl*e. Therefore, while polariton nonlinearities are intrinsically due to molecular anharmonicity, the electromagnetic coherence fundamentally affects the character of these nonlinearities by allowing them to become extended across the size of the optical cavity, and give the striking size-dependent effects observed by our experiments.

## S3.2 Feynman Diagrams for the AC Signal

In Fig. S9, we show the *dominant* Feynman pathways(*22*) (diagrams) that generate the polariton four-wave-mixing (AC) signals (as measured by our experiments) oscillating with the Rabi frequency as a function of delay time $t_2$ (see Sec. S2.4). Each row on the l.h.s of Fig. S9 is labeled by a diagonal or cross-peak in the 2D IR spectrum schematically represented on the r.h.s. To each peak in the AC part of the 2D spectrum, there exists two main diagrams associated to probe-induced stimulated emission (SE) and excited-state absorption (ESA). In the SE pathways, the probe pulse stimulates polariton emission, and the final state is the ground $|g\rangle$. On the other hand, ESA induces transitions from polariton states $|LP\rangle$ or $|UP\rangle$ into two-body excitations $|LP, UP\rangle$. The amplitudes of the ESA and SE pathways have opposite sign. In harmonic systems, these amplitudes also have the same intensity, and therefore, there is no nonlinear signal. The nonlinear AC response measured in our experiment is an indication of the lack of cancellation between the ESA and SE pathways represented in Fig. S9. Physically, as discussed in the main text, a nonlinear dephasing mechanism operates on the $|LP, UP\rangle$ state which prevents destructive interference between the amplitudes of the SE and ESA pathways, ultimately giving rise to finite nonlinear AC signals which last for $4 - 5$ ps.

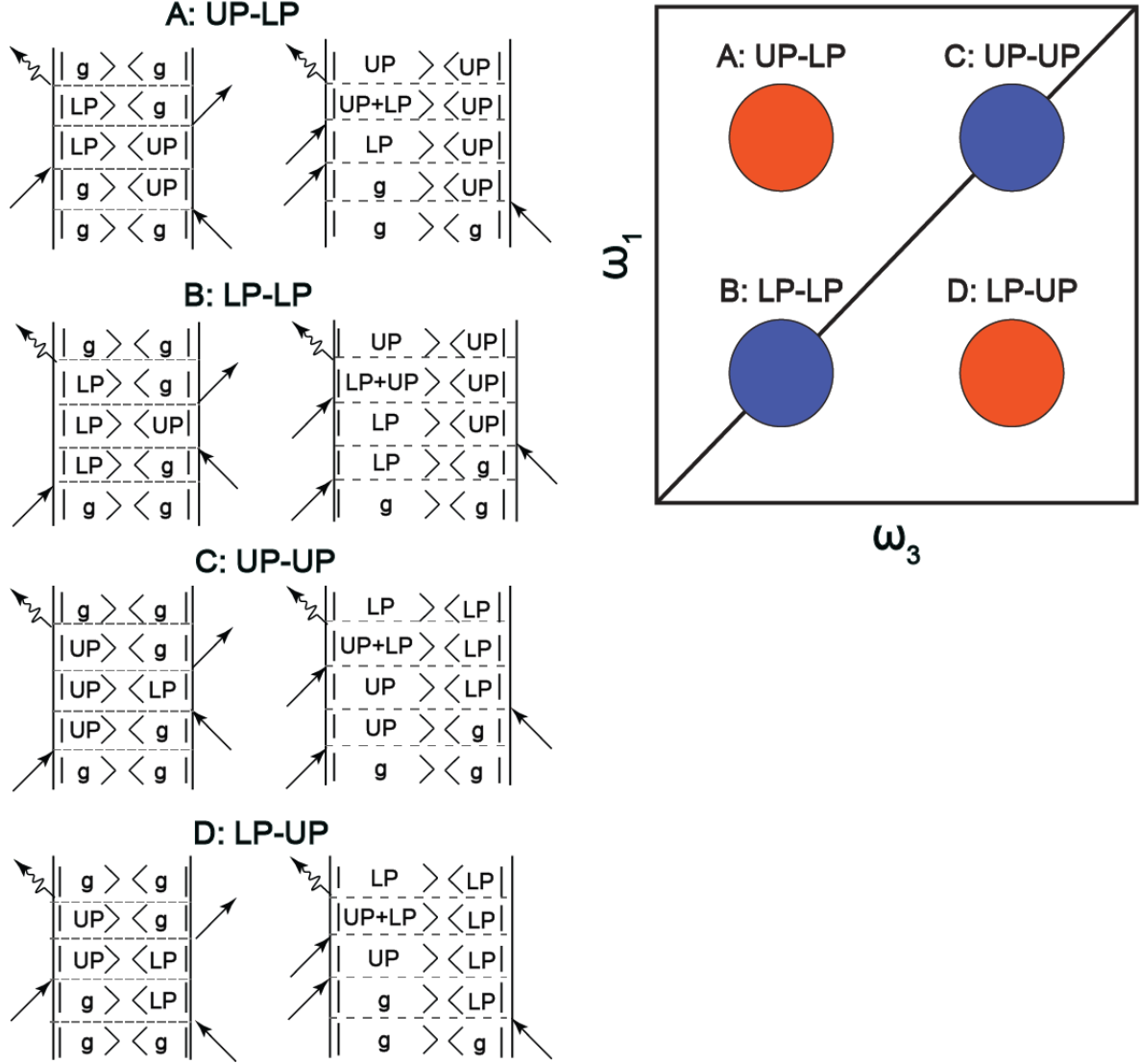

*Figure S9. Feynman diagrams(22) representing the oscillating nonlinear responses (AC components) in each region: A) $\omega_1 = \omega_{UP}$ and $\omega_3 = \omega_{LP}$; B) $\omega_1 = \omega_{LP}$ and $\omega_3 = \omega_{LP}$; C) $\omega_1 = \omega_{UP}$ and $\omega_3 = \omega_{UP}$; D) $\omega_1 = \omega_{LP}$ and $\omega_3 = \omega_{UP}$.*

### S3.3 Phenomenological Simulation of Polariton Bleach

The classical expression for the steady-state transmission intensity of a Fabry-Perot (FP) cavity containing an isotropic, homogeneous, linear absorptive medium characterized by a frequency-dependent linear susceptibility is given by(*1, 32, 45*)

$$T_c(\nu) = \frac{T^2 e^{-\alpha(\nu)L/\cos(\theta)}}{1+R^2 e^{-2\alpha(\nu)L/\cos(\theta)}-2Re^{-\alpha(\nu)L/\cos(\theta)}\cos[4\pi n(\nu)\cos(\theta)\nu]} \quad (12)$$

where ν is the probed frequency, θ = 0.1 rad is the incidence angle, R = 0.94 and T = 0.06 are the reflectance and transmittance of the employed FP cavity mirrors around the fundamental resonance frequency of the absorptive medium $\nu_1$ =1983 cm-1, *L* is the cavity longitudinal length (in the performed simulations, L =

0.00025 cm or 0.000125 cm), $n(\nu)$ is the real part of the cavity mirror complex refractive index and $\alpha(\nu) = 4\pi j(\nu)\nu$, where $j(\nu)$ is the imaginary part of the complex refractive index. As mentioned in the main text, by changing the homogeneous linewidth of fundamental mode of W(CO)$_6$, ($\Gamma_1$), the real and imaginary parts of the refractive index *(n, j)* are modulated accordingly (Eqns. (12) − (14)(*1*)). This effect causes variation of the transmission intensity ($T_c$). Specifically, a larger molecular homogeneous linewidth implies enhanced polariton broadening. The oscillator strength of the molecular vibrations is denoted A$_1$. This quantity is proportional to the density of molecular absorbers. Adjusting A$_1$ does *not* lead to the purely absorptive polariton bleach feature, instead, it causes polariton resonance shifts in frequency, i.e. Rabi splitting contraction. Thus, the semiclassical simulation results indicate that changes of $\Gamma_1$, but not A$_1$, cause polariton bleach features. The following are the quantities which are required for the computation of the transmission with Eq. (12):

$$n(\nu) = \sqrt{\frac{\varepsilon_1 + \sqrt{\varepsilon_1^2 + \varepsilon_2^2}}{2}}, \quad j(\nu) = \sqrt{\frac{-\varepsilon_1 + \sqrt{\varepsilon_1^2 + \varepsilon_2^2}}{2}} \tag{13}$$

where $\varepsilon_1$ and $\varepsilon_2$ are the real and imaginary parts of dielectric constant, expressed as

$$\varepsilon_1 = \varepsilon_{inf} + \sum_{i=1}^{2}\left[\frac{A_i(\nu_i^2 - \nu^2)}{(\nu_i^2 - \nu^2)^2 + (\Gamma_i\nu)^2}\right], \quad \varepsilon_2 = \sum_{i=1}^{2}\left[\frac{A_i\Gamma_i\nu}{(\nu_i^2 - \nu^2)^2 + (\Gamma_i\nu)^2}\right] \tag{14}$$

where we set the background dielectric constant at infinite frequency to be $\varepsilon_{inf} = 2.0135$, $\nu_i$ are the frequencies of the 0 → 1 and 1 → 2 asymmetric stretch transitions of W(CO)$_6$ given by $\nu_1$ = 1983 cm$_{-1}$, and $\nu_2$ = 1968 cm$_{-1}$, and the $\Gamma_i$ are the linewidths of the corresponding vibrational modes ($\Gamma_1$ and $\Gamma_2$ are 3.0 and 4.5 cm$_{-1}$, respectively). The oscillator strength A$_1$ = 3200 cm$_{-1}$ is chosen so the linear transmission resonance frequencies match the experimentally observed with incidence angle θ = 0.0873 rad, while A$_2$ is neglected, since the fundamental mode has the largest population and is in resonance with the cavity photon.

## S3.4 Cavity Coherence Volume

We estimate the coherence volume of the optical cavity by using the following equation(*46*)

$$V_{\text{eff}} = S_{\text{eff}} \cdot l = \frac{\pi l^2 \lambda_\nu}{1-R} \tag{15}$$

where $V_{\text{eff}}$ and $S_{\text{eff}}$ are the effective volume and area of a specific cavity mode, $l$ is the cavity longitudinal length (12 and 25 μm), $\lambda_\nu$ is the wavelength of vibrational transition (5042.86 nm in our case, corresponding to 1983 cm$_{-1}$) and $R$ is the reflectivity of the DBRs which is 96% in the experiments.

Using Eq. (7), we find the coherence volumes of the 12 and 25 μm systems are 0.57×10$_5$ and 2.5×10$_5$μm$_3$, respectively, i.e., the cavity with longest length has a coherence volume that is approximately four times greater than that with the shortest.